\documentclass[aps,prl,reprint,noaffiliation,groupedaddress,amsmath,amssymb,]{revtex4-2}
\usepackage{graphicx}%
\usepackage{float}
\usepackage[x11names]{xcolor}
\usepackage[colorlinks,citecolor=blue]{hyperref}

\begin{document}
	
    \title{Emerging Multidimensional Real-Space Topological Structures at \\ Chiral Bound States in the Continuum}

	\author{Xingqi Zhao$^{1}$}
    \author{Jingguang Chen$^{1}$}
	\author{Jiajun Wang$^{1,5}$}
	\email{jiajunwang@fudan.edu.cn}
    \author{Lixi Rao$^{1}$}
	\author{Wenzhe Liu$^{1,2,5}$}
    \author{Fang Guan$^{1,2,5}$}
    \author{C. T. Chan$^{3}$}
	\author{Lei Shi$^{1,2,4,5}$}
    \email{lshi@fudan.edu.cn}
	\author{Jian Zi$^{1,2,4,5}$}
    \email{jzi@fudan.edu.cn}
	\affiliation{$^{1}$State Key Laboratory of Surface Physics, Key Laboratory of Micro- and Nano-Photonic Structures
		(Ministry of Education) and Department of Physics, Fudan
		University, Shanghai 200433, China}
	\affiliation{$^{2}$Institute for Nanoelectronic devices and Quantum computing, Fudan University, Shanghai 200438, China}
    \affiliation{$^{3}$Department of Physics, The Hong Kong University of Science and Technology, Clear Water Bay, Hong Kong, 999077, China}
	\affiliation{$^{4}$Collaborative Innovation Center of Advanced Microstructures, Nanjing University, Nanjing 210093, China}
	\affiliation{$^{5}$Shanghai Research Center for Quantum Sciences, Shanghai 201315, China}



	
\begin{abstract}
As widely studied topological singularities, bound states in the continuum (BICs) have revealed rich physical properties through their momentum-space topology. Here, we reveal and experimentally demonstrate that magnetically induced chiral BICs possess multidimensional topological structures extending into real space. We design and realize a gyromagnetic photonic crystal slab where magnetic field breaks the time-reversal symmetry and lifts the degeneracy of BICs, creating 
a pair of chiral BICs with opposite circular polarizations. Near-field scanning 
measurements reveal phase vortices with quantized topological charges, spatially distributed near-field chirality, and skyrmionic Stokes 
textures arising from magnetic control. Our work unveils a previously unexplored 
dimension of BIC topology and establishes gyromagnetic photonic crystals as versatile 
platforms for manipulating complex topological states.
\end{abstract}

\maketitle

Topological structures - from phase vortices \cite{gbur2016singular,quinteiro2022interplay} and topological defects \cite{mermin1979topological} to more exotic configurations like skyrmions \cite{nagaosa2013topological,shen2024optical} - appear across various physical systems and exhibit distinct topological properties. These structures not only provide frameworks for understanding physical phenomena but also exhibit rich physical effects themselves \cite{blatter1994vortices,abo2001observation,fert2017magnetic,ni2021multidimensional}. For instance, magnetic flux quantization in superconductors are connected with topological vortices \cite{blatter1994vortices}; in classical wave systems, vortex fields can govern energy flow and create localized field enhancement \cite{wang2025topological,zhao2025topological}. The rich physics and effects of topological structures have motivated their exploration across various research fields.

In modern photonics and optics, the explorations of topological structures have proven particularly fruitful. These singular optical fields, manifesting in various forms from structured polarization configurations to phase vortices and optical skyrmions \cite{gbur2016singular,zhan2009cylindrical,he2022towards,chen2022multidimensional,shen2024optical}, provide powerful capabilities and new degrees of freedom in light manipulation \cite{kleckner2013creation,bauer2015observation,hu2025topological}, communications \cite{willner2015optical,willner2021orbital}, information processing \cite{bozinovic2013terabit,ndagano2017characterizing}, sensing \cite{lavery2013detection}, and light-matter interaction \cite{he1995direct,quinteiro2022interplay}. To explore and harness these topological optical fields, various optical platforms have been developed. Among them, engineered photonic structures like photonic crystals (PhCs) have created a diverse landscape of topological physics \cite{lu2014topological,ozawa2019topological}. The interaction between light and periodic structures provides a versatile platform for exploring photonic topological structures in multiple dimensions, leading to interesting discoveries and realizations such as topological valleys \cite{dong2017valley}, Dirac vortex \cite{gao2020dirac}, vortex beam \cite{wang2020generating}, and spatiotemporal vortex \cite{zhang2025bulk}.

Bound states in the continuum (BICs), emerging as topological singularities in momentum space, represent a particularly intriguing class of topological states in PhC slabs \cite{hsu2016bound,huang2023resonant,kang2023applications,wang2024optical}. Studies have established that BICs exhibit rich topological characteristics: they carry polarization vortices in momentum space, which can lead to the generation of vectorial beams, phase vortices, and even skyrmionic textures \cite{huang2020ultrafast,wang2020generating,wang2025inherent,rao2025meron}. To date, BICs are primarily explored by their momentum-space topology \cite{zhen2014topological,zhang2018observation,Doeleman2018experimental,liu2019circularly}, even though real-space electromagnetic field distributions fundamentally reflect elaborate characteristics of these singular states and directly mediate light-matter interactions. The real-space topological nature of BICs, such as the spatial topological configurations of electromagnetic fields, remains largely unexplored. Recent theoretical advances have revealed the emergence of phase vortices in the near-field distribution of chiral BICs \cite{zhao2024spin}, suggesting that momentum-space topology connects unexplored real-space field configurations. It inspires a broad landscape for exploring multidimensional topological structures across momentum and real space through BICs, opening new avenues in topological photonics.

In this work, we reveal multidimensional real-space topological structures in magnetically induced chiral BICs. By applying an 
magnetic field to a gyromagnetic PhC slab composed of yttrium iron garnet (YIG) 
cylinders, we 
transform doubly degenerate BICs 
into a pair of chiral BICs with opposite handedness in the far field. Through 
near-field 
measurements, we uncover that these chiral BICs host 
multidimensional
real-space topological structures: phase vortices,
spatially distributed near-field 
chirality patterns, and skyrmionic Stokes 
textures. All these topological structures can be switched by reversing the magnetic field. Our findings establish a new paradigm 
connecting momentum-space and real-space topology in BICs and open new avenues 
for magnetically tunable topological structures.

\begin{figure}[htp]
    \centering
    \includegraphics[scale=1]{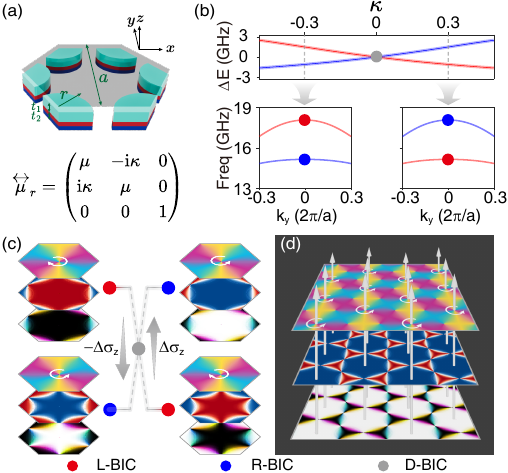}
    \caption{(a) 
    Upper panel: a unit cell of the PhC slab, where the gray plane indicates where the near-field distributions are evaluated in later discussions. 
    Lower panel: the relative permeability tensor $\overset{\leftrightarrow}{\mu}_r$ of YIG. (b) Top: evolution of the BICs with $\kappa$. 
    Bottom: representative photonic band structures for 
    $\kappa=\pm 0.3$. Red/blue/gray dots denote L-BIC/R-BIC/D-BIC, respectively.  (c) Real-space topological textures in a unit cell associated with the chiral BICs, evaluated at the gray plane shown in (a). From top to bottom: (i) phase distributions, (ii) chirality distributions, and (iii) Stokes vector distributions. Reversing the magnetic field swaps the two chiral BICs and their associated topological textures. (d) Schematic real-space topological structures of the R-BIC across the PhC slab.}
    \label{Fig1}
\end{figure}


Figure \ref{Fig1}(a) shows the schematic view of a unit cell of the gyromagnetic PhC slab, which comprises a honeycomb lattice of YIG cylinders.  Here, the radius $r$ and thickness $t_1$ of the YIG cylinder are $3$ mm and $1.4$ mm, respectively. The lattice constant $a$ is $14$ mm. In the absence of a magnetic field, the structure supports a pair of doubly degenerate BICs (D-BICs) at the $\Gamma$ point. When the YIG cylinders are magnetized by permanent magnets placed beneath the slab, where the magnets have the thickness $t_2=1$ mm and the same radius $r$, the YIG exhibits a gyromagnetic response described by the relative permeability tensor \cite{pozar2021microwave}

\begin{equation}
    \overset{\leftrightarrow}{\mu}_r=\begin{pmatrix}
        \mu & -\mathrm{i}\kappa & 0  \\
        \mathrm{i}\kappa & \mu & 0\\
        0 & 0 & 1
    \end{pmatrix}.
\end{equation}
The off-diagonal element $\kappa$ is controlled by the out-of-plane magnetic field. With this magnetic field, the degeneracy is lifted. At the $\Gamma$ point, the coupled‑mode dynamics can be captured by the effective Hamiltonian \cite{Wang2005magneto}:
\begin{equation}
    \mathrm{H}=E_0+\Delta \sigma_z.
\end{equation}
Here $E_0$ is the energy of the degenerate pair, and $\Delta$ is the Zeeman term which is governed by the magnetic field (the value of $\kappa$), controlling the splitting of modes. As shown in the upper panel of Fig. \ref{Fig1}(b), increasing $|\kappa|$ continuously separates the eigenfrequencies of modes with an energy gap $\Delta E$, and two chiral BICs emerge from the original D-BICs. Two chiral BICs carry opposite handedness: In the vicinity of them, the radiative channels exhibit circular polarization in far field, one corresponds to left-hand circular polarization (L-BIC) and the other to right-hand circular polarization (R-BIC) \cite{zhao2024spin}. Flipping the magnetic field flips the sign of $\kappa$ and swaps the handedness of the two branches, as shown in Fig. \ref{Fig1}(b).

The topology of BICs is typically characterized in momentum space through far-field radiation properties \cite{wang2024optical,kang2023applications}. Here, we reveal that these chiral BICs exhibit unprecedented richness in real-space topology, manifested in their near-field distributions. Figure \ref{Fig1}(c) exhibits the near-field topological structures in a unit cell for the paired chiral BICs under the two opposite magnetization directions described in Fig. \ref{Fig1}(b), showing topological characteristics that extend beyond the conventional far-field properties of BICs. The six-fold rotational ($C_6$) symmetry of the structure enforces a phase vortex at the unit cell center. From group theory, the paired chiral BICs correspond to the $E_{2a}$ and $E_{2b}$ representations of the $C_6$ point group, which require the complex field to acquire a winding phase of $\pm 4\pi$ around the rotation center for the R-BIC (L-BIC).  This phase vortex is accompanied by spatially distributed chirality in the near field. The near field chirality is made explicit by the Stokes parameter $S_3/S_0$, with $S_3>0$ corresponds to right-handed circularly polarized (RCP) and $S_3<0$ corresponds to left-handed circularly polarized (LCP). There is a pronounced difference of chirality between the vicinity of the YIG region and the surrounding areas, revealing a strongly localized contrast.  When considering the full Stokes vector field, it reveals that the polarization further organizes into a topological skyrmionic texture in real space. Note that, upon reversing the magnetization, all these topological structures reverse accordingly. When considering the whole PhC slabs (Fig. \ref{Fig1}(d)), they form a coherent periodic pattern, highlighting that the topological characteristics of chiral BICs extend beyond momentum space to robust real-space configurations. 

\begin{figure}[h]
    \centering
    \includegraphics[scale=1]{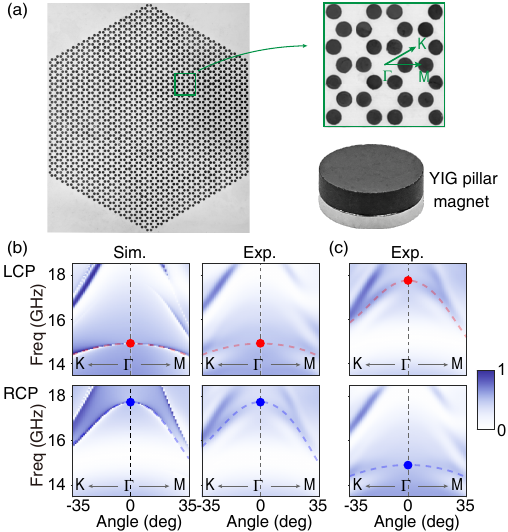}
    \caption{(a) Photo of the experimental sample. 
    (b) Simulated and measured 
    co-polarized transmission spectra under circularly polarized incidence for the sample magnetized in the positive B direction. Upper row: LCP incidence; Lower row: RCP incidence. Dashed curves mark bands whose handedness matches that of the incident light. (c) Measured co-polarized transmission spectra under reversed magnetization.}
    \label{Fig2}
\end{figure}

To experimentally realize the chiral BICs, we fabricated the designed PhC slab, with a photo shown in Fig. \ref{Fig2}(a). Each YIG pillar is mounted on a magnet, and the YIG–magnet units are embedded in a foam slab perforated with a honeycomb lattice, as shown in the right panel. With the external magnetic field, two bands carry opposite signs of the $S_3$ in far field, indicating opposite handedness of their radiation. The calculated band structures are provided in the Supplemental Material Sec. 2.


The opposite chirality leads to distinct transmission responses under circularly polarized incidence. Figure \ref{Fig2}(b) presents the co‑polarized transmission spectra for positive B, where the magnetic field is defined as positive when the north pole of the magnets upward. In this configuration, modes in the lower band radiate predominantly LCP, and those in the upper band radiate predominantly RCP. Consequently, near $\Gamma$ point, LCP incidence selectively excites the lower band (traced by the red dashed curve), whereas the upper band is not efficiently excited and thus barely visible in the spectrum. Under RCP incidence, the upper band is excited (blue dashed curve) while the lower band becomes inconspicuous for the same reason. 
Note that the $\Gamma$ points of the two bands correspond to BICs, which cannot be excited by any external incidence and thus do not appear in the transmission spectra for either polarization. When reversing the magnetic field, the chiral responses swap (Fig. \ref{Fig2}(c)). These transmission measurements directly demonstrate the realization of chiral BICs from the far field, and characterize their magnetically controllable chiral responses with the far-field incidence.

\begin{figure}[h]
    \centering
    \includegraphics[scale=1]{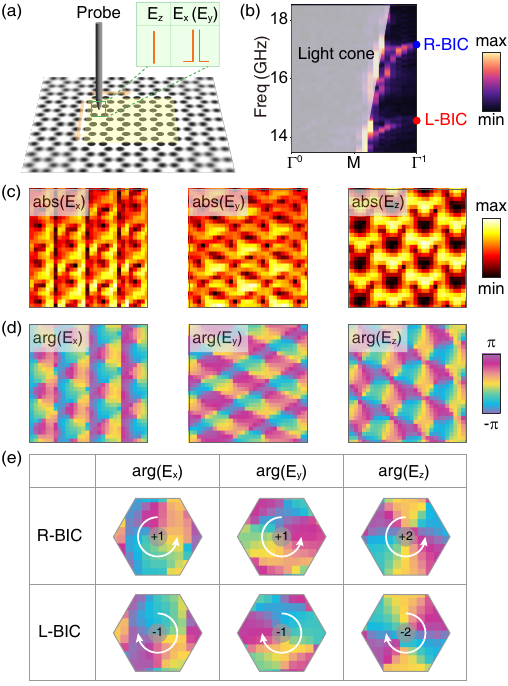}
    \caption{(a) Schematic view of the near-field scanning setup. 
    (b) The measured near-field Fourier components of the $E_z$ amplitude, showing the two bands of interest. The gray region denotes the light cone.  (c)-(d) Measured  amplitude (c) and phase (d) of the electric field for the R-BIC. (e) The phase vortices in the unit cell for the R-BIC and L-BIC, where the L-BIC has opposite phase winding compared to the R-BIC.}
    \label{Fig3}
\end{figure}

We then characterize the near-field properties of the chiral BICs. Figure \ref{Fig3}(a) shows a schematic of the near-field scanning setup: a probe scans above the PhC surface to map the spatial distribution of the electric field. Two types of probes are used for the measurement of normal electric field ($E_z$) and tangential electric field ($E_x$ and $E_y$). Here, the measured Fourier spectrum of the $E_z$ component (Fig. \ref{Fig3}(b)) reveals the two bands of interest. By selecting Fourier components in the vicinity of the chiral BICs and performing inverse Fourier transform, we can extract the real-space near-field distribution of chiral BICs. Details of the experimental setup and data analyze are discussed in Supplemental Material Sec. 7.

Figures \ref{Fig3}(c) and \ref{Fig3}(d) present the measured amplitude and phase distributions of the electric field for the R-BIC. Notably, all field components vanish at the unit-cell center, manifesting as zero-intensity points. The phase distributions reveal that these zero-amplitude points are phase singularities: the phase winds continuously around the center, forming vortices in real space. The topological charge $q$ of these phase vortices is defined as \cite{hsu2016bound}
\begin{equation}
    q=\frac{1}{2\pi}\oint_{C} \mathrm{d}\boldsymbol{r} \cdot \nabla\phi (\boldsymbol{r}),
\end{equation}
where $\phi$ is the phase, $\boldsymbol{r}$ is the spatial coordinate, and the integration is performed along a closed loop $C$ around the singularity. Figure \ref{Fig3}(e) summarizes the topological phase structures in a unit cell. For the R-BIC, all field components ($E_x$, $E_y$, $E_z$) exhibit counterclockwise phase winding around the center, with $E_x$ and $E_y$ carrying a topological charge of $q = +1$, and $E_z$ carrying $q = +2$. In contrast, the L-BIC exhibits opposite phase winding in all components. This reversal of vortex winding between the paired chiral BICs demonstrates that the photonic pseudospin (corresponding to chirality) is intrinsically locked to the orbital angular momentum (corresponding to the phase vortices) in this system \cite{zhao2024spin}.

\begin{figure}[h]
    \centering
    \includegraphics[scale=1]{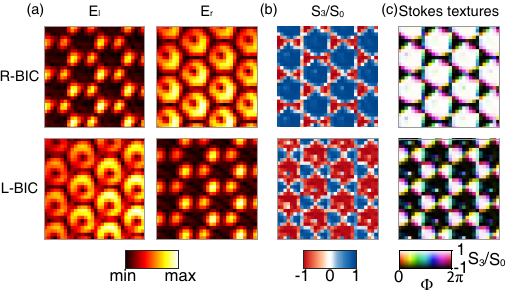}
    \caption{(a) The left-handed circular (left panel) and right-handed circular (right panel) components of the near-field electric field of chiral BICs. (b) The normalized third Stokes parameter ($S_3/S_0$) of chiral BICs in the near field. (c) The distribution of Stokes vectors, showing skyrmionic textures. From top to bottom: measured results of R-BIC and L-BIC.}
    \label{Fig4}
\end{figure}

We further investigate the near-field chirality of the chiral BICs. In Fig. \ref{Fig4}(a), we provide the left-handed and right-handed circular components ($E_l$ and $E_r$) of the electric field, which are defined by
\begin{equation}
E_{l(r)}=E_x\pm\mathrm{i}E_y.
\end{equation}
For each BIC, the near field exhibits a distinct chirality distribution: For the R-BIC, the $E_l$ component is mainly localized around the YIG cylinders, while the $E_r$ component is mainly localized in other regions. Note that the center of the unit cell is a singularity with zero amplitude for both $E_l$ and $E_r$. This spatially separated chirality distribution reveals that the magnetic field not only induces far-field chirality in the radiation, but also creates unique chiral textures in the near field, with opposite circular polarizations confined to distinct spatial regions. The L-BIC shows the opposite pattern, with $E_l$  and $E_r$ distributions reversed compared to the R-BIC. This chirality distribution is quantitatively captured by the normalized third Stokes parameter $S_3/S_0$, which measures the degree of circular polarization. As shown in Fig. \ref{Fig4}(b), the $S_3/S_0$ map reveals strong spatial variation. For the R-BIC, negative values are concentrated around the YIG cylinders and positive values in other regions, confirming the spatially segregated chirality distribution observed in Fig. \ref{Fig4}(a). For the L-BIC, the sign distribution of $S_3/S_0$ is reversed. Note that the near-field chirality here is characterized by the circular polarization of the in-plane electric field \cite{lodahl2017chiral,sollner2015deterministic,shreiner2022electrically,chen2022multidimensional}, which differs from the optical chirality density involving both electric and magnetic fields \cite{chen2022multidimensional}.

Nontrivial topological distributions in vector fields represent fundamental properties with inherent stability, exemplified by skyrmionic textures, which have recently garnered considerable interest across diverse physical systems \cite{nagaosa2013topological,fert2017magnetic,shen2024optical,Kara2025multistable,Ma2025nanophotonic,Huh2025stable,Schwab2025skyrmion}. Here we reveal that similar topological textures also appear in the real-space 
near-field distribution for chiral BICs. Considering the full normalized Stokes vector $\boldsymbol{S} = (S_1,S_2,S_3)/S_0 $ of the chiral BIC, we observe skyrmionic organization in real space, as shown in Fig. \ref{Fig4}(c). Two key features characterize these textures: 
(1) the $S_3$ component changes sign from the center to the boundary of the unit 
cell, representing a transition from one circular polarization to the opposite; 
(2) the azimuthal angle $\Phi = \mathrm{arg}(S_1+\mathrm{i}S_2)$ winds continuously 
around the center. Notably, the paired R-BIC and L-BIC exhibit opposite signs 
of $S_3$ while maintaining the same winding direction of $\Phi$. These real-space 
near-field Stokes textures demonstrate that complex topological structures are also
intrinsic properties of optical modes in PhC slabs, embodying rich polarization configurations in the confined electromagnetic fields.

\begin{figure}[h]
    \centering
    \includegraphics[scale=1]{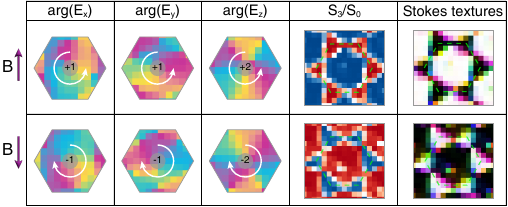}
    \caption{ 
    Upon reversing the magnetic field, the BIC on the upper band transforms from R-BIC to L-BIC, with its near-field topological structures switching correspondingly.}
    \label{Fig5}
\end{figure}

The real-space topological structures are governed by the direction of the magnetic field. To experimentally verify and demonstrate the magnetic control, we fabricated another sample with the magnets oriented oppositely. This reverses the magnetic field
, flips the sign of $\kappa$, and swaps the handedness of the two chiral BICs. The upper band, which supports an R-BIC in the first sample, now hosts an L-BIC; conversely, the lower band switches from L-BIC to R-BIC.

Figure \ref{Fig5} compares the measured near-field topological structures of the upper-band BIC for both magnetization directions. Under upward magnetization (top row), the BIC exhibits counterclockwise phase winding in all the electric field components, negative $S_3$ in the YIG regions, and corresponding Stokes vector textures. Upon reversing the magnetic field (bottom row), all these topological features reverse: the phase vortices wind clockwise, the chirality distribution inverts, and the Stokes textures flip accordingly. These experimental results confirm that the real-space topological structures are intimately linked with the chiral BICs and can be deterministically switched by the magnetic field direction.

In conclusion, our work unveils rich real-space topological structures at chiral BICs beyond conventional momentum-space characterizations.
Applying an external magnetic field to a gyromagnetic PhC slab not only creates chiral BICs with opposite handedness in the far field but also generates multidimensional real-space topological structures in the near field, including phase vortices, spatially
distributed chirality, and skyrmionic Stokes textures. These topological structures can enhance nonlinear processes and sensing through chiral field localization, and create new paradigms for chiral light-matter interactions and topological phase engineering. 
This magnetic control of real-space topological structures establishes a new paradigm for topological photonics, enabling on-demand manipulation of topological states. Future directions include exploring magnetic field control of other topological singularities beyond degenerate BICs, investigating additional topological invariants in these systems, and studying other 
topological structures. These investigations may reveal new topological phenomena and advance our fundamental understanding of multidimensional topology in photonic systems.
	
\bigskip
	
\begin{acknowledgments}
This work is supported by National Key R\&D Program of China (No. 2023YFA1406900, No. 2022YFA1404800, and No. 2021YFA1400603); National Natural Science Foundation of China (No. 12404427,
 No. 12234007, No. 124B2084, No. 12321161645, No. 12221004, No. 125B2087, No. T2394480, and No. T2394481); Science and Technology Commission of Shanghai Municipality (No. 24YF2702400,
 No. 2019SHZDZX01, 
 No. 23DZ2260100, and
 No. 24142200100);  Research Grants Council (RGC) of Hong Kong (CRS\_HKUST601/23).
		
\bigskip

X.Z., J.C. and J.W. contributed equally to this Letter.
		
\end{acknowledgments}

\end{document}